\documentclass[10pt,peerreview,draftcls,onecolumn]{IEEEtran}

\usepackage{graphicx}
\usepackage{multirow}
\usepackage{subfigure}
\usepackage{url}
\usepackage{epstopdf}
\usepackage{color}
\usepackage{xcolor}
\usepackage{bbm}
\usepackage[cmex10]{amsmath}
\usepackage{amssymb}
\usepackage{array}
\usepackage{caption}
\colorlet{mygreen}{green!60!gray}

\newcommand{\BT}[1]{\textcolor{blue}{{#1}}}

\begin{document}
\title{An Improved Statistic for the Pooled Triangle Test against PRNU-Copy Attack} %in \\ PRNU-based Camera Identification}

\author{*Mauro Barni,~\IEEEmembership{Fellow,~IEEE}, H{\'e}ctor Santoyo Garc{\'i}a, Benedetta Tondi,~\IEEEmembership{Member,~IEEE}%
\thanks{M. Barni, B. Tondi are with Dept. of Information Engineering and Mathematical Sciences, University of Siena, Italy (barni@dii.unisi.it, benedettatondi@gmail.com). H. Santoyo-Garc{\'i}a is with the Postgraduate Section, Mechanical Electrical Engineering School, National Polytechnic Institute of Mexico, Mexico (hectorsantoyogarcia@gmail.com).  \newline
This work has been partially supported by a research sponsored by DARPA and Air Force Research Laboratory (AFRL) under agreement number FA8750-16-2-0173. The U.S. Government is authorised to reproduce and distribute reprints for Governmental purposes notwithstanding any copyright notation thereon. The views and conclusions contained herein are those of the authors and should not be interpreted as necessarily representing the official policies or endorsements, either expressed or implied, of DARPA and Air Force Research Laboratory (AFRL) or the U.S. Government.
\newline * The list of authors is provided in alphabetic order.}
\vspace{-0.5cm}
}

%% The paper headers
%\markboth{IEEE SIGNAL PROCESSING LETTERS, VOL. X, NO. X, MONTH 20XX}%
%{Massai \MakeLowercase{\textit{et al.}}: Bare Demo of IEEEtran.cls for Journals}
%% The only time the second header will appear is for the odd numbered pages
%% after the title page when using the twoside option.
%%
%% *** Note that you probably will NOT want to include the author's ***
%% *** name in the headers of peer review papers.                   ***
%% You can use \ifCLASSOPTIONpeerreview for conditional compilation here if
% you desire
\maketitle

\begin{abstract}
We propose a new statistic to improve the pooled version of the triangle test used to combat the fingerprint-copy counter-forensic attack against PRNU-based camera identification \cite{Goljan}. As opposed to the original version of the test, the new statistic exploits the one-tail nature of the test,  weighting differently positive and negative deviations from the expected value of the correlation between the image under analysis and the candidate images, i.e., those image suspected to have been used during the attack. The experimental results confirm the superior performance of the new test, especially when the conditions of the test are challenging ones, that is when the number of images used for the fingerprint-copy attack is large and the size of the image under test is small.
\end{abstract}


\IEEEpeerreviewmaketitle

\enlargethispage{\baselineskip}

\section{Introduction}
\label{sec.intro}

Photo-Response Non Uniformity (PRNU) noise \cite{PRNU} has been successfully used for forensic camera identification  \cite{cameraIdent} and image forgery detection \cite{BayesImgFor,lukavs2006detecting}. Techniques based on PRNU are prone to the so-called fingerprint-copy  (or PRNU-copy) attack \cite{Bohme}, according to which, a forger, usually referred to as Eve, estimates the PRNU from a set of publicly available images acquired by the camera of a victim, say Alice, and implant the estimated PRNU into an image shot by a different camera.
An effective countermeasure against the fingerprint-copy attack is the {\em triangle test} proposed in \cite{Goljan}.
The test exploits the fact that an image forged with the fingerprint-copy attack shares with the images used by Eve to estimate the PRNU other noise components in addition to the PRNU, hence resulting in an unnaturally high correlation between the forged image and the images used to create the forgery. In its simplest version, the triangle test allows Alice to understand {\em which} images, in a set of publicly available images acquired by her camera, have been used to produce the forgery.
In other cases, Alice's goal is {\em just} to prove that the image under analysis has been forged by means of a fingerprint-copy attack, without the need to identify the exact subset of images used to produce the forgery. To do so, Alice can resort to the {\em pooled} version of the test \cite{Goljan}. The pooled test is generally very powerful and
the effectiveness of the counter-forensic methods proposed so far against the single-image triangle test, e.g. \cite{rao2013anti,napoli,Caldelli}, is dramatically reduced when the pooled triangle test is considered.

In this paper, we propose a refined statistic for the pooled triangle test, that allows to improve the performance of the test with particular reference to those situations where the test is less reliable, namely when the number of images Eve has access to is large and when the size of the analysed image is small. The improved statistic relies on the observation that the original pooled test treats in the same way both images exhibiting an unnaturally high correlation with the image under test and those for which this correlation is lower than expected. In this way, the analysis somewhat neglects the one-tail nature of the test\footnote{We remark that such an observation does not apply to the single-image version of the test (see eq. (16) in \cite{Goljan}).} according to which the images used for the PRNU-copy attack are expected to exhibit a larger correlation with respect to those that have not been used to create the forgery. The new statistic, on the contrary, accumulates the deviations from the expected correlation by considering their sign. The resulting test, then, decides that the image under analysis has been subject to a PRNU-copy attack only in the presence of positive deviations.
The superior performance of the proposed statistic are assessed in a wide variety of cases, by varying the parameters that impact most on the performance of the test, that is, the number $N$ of images used by Eve to estimate the PRNU, the overall number $N_c$ of public images available, and the size of the images.

%In order to better assess the reliability of the modified pooled test, we considered two possible versions of the test, corresponding to two different interpretations of the false alarm error probability. Our experiments show that the pooled test based on the new statistic is more reliable than the original version, especially when the number of images $N$ used by Eve is large, and, hence, Eve's estimation of the PRNU fingerprint is a very good one.

The paper is organized as follows. In Section \ref{sec.brief_description}, we review the PRNU-copy attack and the pooled triangle test. The proposed improved statistic is described in Section \ref{sec.proposed_test}. The results of the experimental validation  are presented and thoroughly discussed in Section \ref{sec.exp}. Eventually, we draw our conclusions and present some directions for future work in Section  \ref{sec.conc}.

\section{PRNU-copy attack and pooled triangle test}
\label{sec.brief_description}

Let us denote with $\mathcal{C}_{1,pub}$ a  public dataset of $N_c$ images acquired by Alice's camera $\textrm{C}_1$. Eve's goal is to take an image $J$ coming from another camera $\textrm{C}_2$ and modify it in such a way that it looks like as if it was generated by $\textrm{C}_1$. To do so, Eve estimates the PRNU of $\textrm{C}_1$ from a subset of $N$ images, $I_i$, $i=1,..,N$, belonging to $\mathcal{C}_{1,pub}$, as follows:
\begin{equation}
\label{estimator}
\hat{K}_E = \frac{\sum_{i=1}^N W_{I_i} I_i}{\sum_{i=1}^N I_i^2},
\end{equation}
where $\hat{K}_E$ is the PRNU estimate obtained by Eve, $W_{I_i} = I_i - F(I_i)$ is the noise residual of $I_i$, and $F$ is a denoising filter, e.g. the one  in \cite{wavelet}.
% \BT{(as in \cite{Goljan}, we consider a wavelet denoising \cite{wavelet})}.
The noise residual has the form  $W_{I_i} = I_i K + \theta$, where $K$ is the true PRNU of $\textrm{C}_1$ and $\theta$ collects the non-PRNU noise components of the residual \cite{PRNU}. Then, Eve superimposes the estimated PRNU onto $J$, obtaining the forged image
%The forged image $J'$ is then obtained as
%
\begin{equation}
\label{forging}
J' = [J(1 + \alpha \hat{K}_E)],
\end{equation}
where $[\cdot]$ indicates rounding to integers and $\alpha$ is the fingerprint strength. The value of $\alpha$ must be sufficiently large to pass the threshold-based correlation test  (see below), but, at the same time, as small as possible to make the forgery undetectable.

On the analyst side, camera attribution is carried out by relying on a threshold-based correlation test, that is by computing $\rho = \text{corr}(W_{I}, I \hat{K}_A)$, where $I$ is the image under test, and $\hat{K}_A$ is Alice's estimation of the PRNU fingerprint of $\textrm{C}_1$, which can be reliably obtained from a limited number of flat-field images.  Image $I$ is attributed to $\textrm{C}_1$, if $\rho$ is above a threshold, set by fixing the false alarm probability.
The forged image $J'$  can easily pass the correlation test \cite{Bohme}, thus being wrongly attributed to $\textrm{C}_1$.

%%%% Mauro %%%
%If the value of $\rho$ is above a detection threshold, usually set by fixing the false alarm probability, the image $I$ is attributed to $\textrm{C}_1$. Previous works (see for instance \cite{Bohme}) have shown that when the number of images in $\mathcal{C}_{1,pub}$ is large enough, the forged image $J'$ can easily pass the correlation test and be wrongly attributed to $\textrm{C}_1$.
%\BT{[La frase sopra secondo me si puo' togliere (sono cose affrontate in [6]). Mi limiterei a 'the forged image $J'$  can
%easily pass the correlation test.']}

As a countermeasure, Alice can apply the {\em triangle test}  \cite{Goljan} to the images attributed to $\textrm{C}_1$, to determine if they are genuine images shot by $\textrm{C}_1$, or they are the result of a PRNU-copy attack. The idea behind the triangle test is the following:  each image $I_i$ used by Eve to estimate $K$, shares with the forged image $J'$ not only the PRNU term (as it happens for a genuine - non forged - image), but also the other terms of the noise residual $W_{I_i}$; then, the correlation of the residual of $J'$ with the one of $I_i$, namely $c_{I_i,J'} = \text{corr}(W_{I_i}, W_{J'})$, is typically larger when $J'$ is a forgery and image $I_i$ has been used to estimate the fingerprint implanted in $J'$.
%the attacked image $J'$ also carries out some information on the images $I_i$ used to estimate the fingerprint in \eqref{{estimator}}.

\begin{figure}
\centering
\includegraphics[width = 0.8\columnwidth]{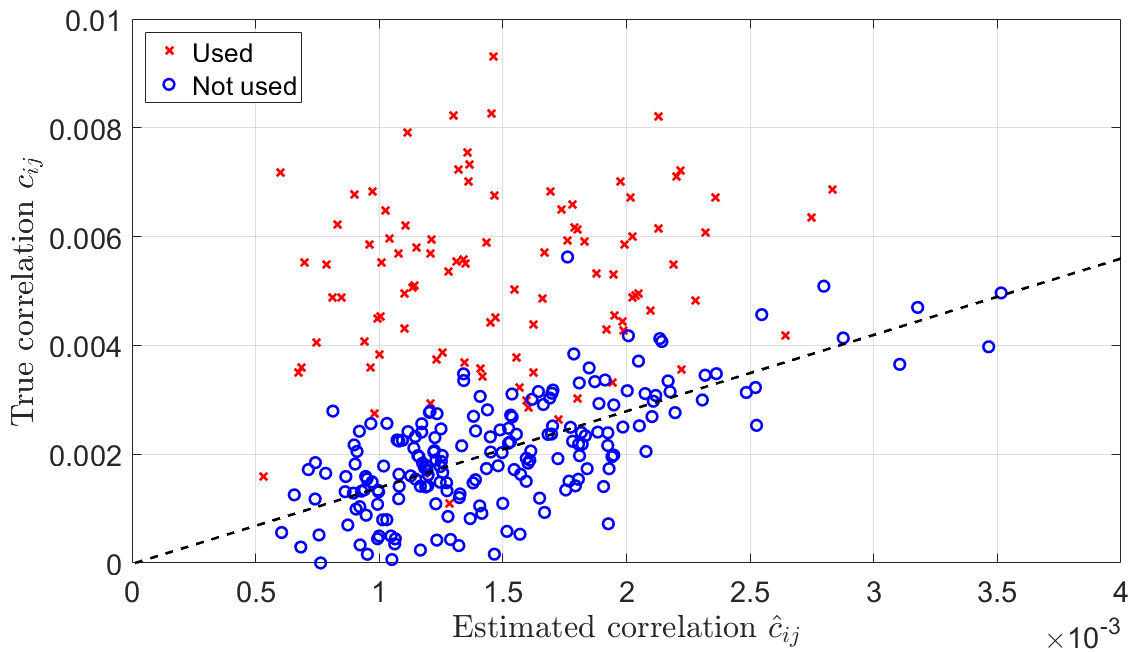} %  0.48\paperwidth
\caption{True correlation ${c}_{I, J'}$  as a function of the estimated correlation $\hat{c}_{I, J'}$  for an image $J'$ forged by Eve with $N = 100$ ($N_c = 300$).}
%\BTdo{Enlarge fonts axis and legends. Remove the title in the top. Enlarge blue points} \MB{The subscripts in the plot should be uppercase. Explain the meaning of red and blues points (clarify the legend). What does $J_{195}$ mean ?}}
%of the idea behind Theorem \ref{theo_attack_H0_H1}. The behavior of the attack channel $A^*_{\Delta}$ is illustrated.}
\vspace{-0.5cm}
\label{fig.triangle_test}
\end{figure}

By following \cite{Goljan}, given a non-forged image $J$ and an image $I$ from $\textrm{C}_1$, it is possible to compute the expected value of $c_{I, J}$, named $\hat{c}_{I, J}$.
%\BT{[this is also, I guess, the expected value when there is no forgery , i.e., $J'$ is not forged]}.
%
The dependence between the real value of $c_{I, J}$ and $\hat{c}_{I, J}$ when $I$ has not been used by Eve to forge $J$, is well fit by a straight line, hereafter referred to as {\em inference line}, $ {c}_{I, J}  = \lambda \hat{c}_{I, J} + \eta$, for some slope $\lambda$ and intercept $\eta$.
On the contrary, if $I$ has been used by Eve to forge $J'$, the correlation $c_{I, J'}$ takes much larger values.  Figure \ref{fig.triangle_test} illustrates a typical plot of $c_{I, J'}$ as a function of $\hat{c}_{I, J'}$ for a forged image $J'$ for $N = 100$, when $N_c = 300$.
% (upper), and in a more difficult case in which $N = 500$, with $N_c = 600$ (lower). The points in red are the correlation values for the $N$ images used by Eve to forge $J'$, those in blue are the values corresponding to the remaining $N_c - N$ images in the candidate set, not used by Eve. }
%Figure \ref{fig.triangle_test}(b) depicts a more difficult case in which $N = 500$ ($N_c = 600$).}
%
%\begin{figure}
%\centering
%\subfigure{\includegraphics[width = 0.78\columnwidth]{plots/prova.png}}\\ %  0.48\paperwidth
%\subfigure{\includegraphics[width = 0.78\columnwidth]{plots/prova2.png}}
%\caption{True correlation ${c}_{I, J'}$  as a function of the estimated correlation $\hat{c}_{I, J'}$  for an image $J'$ forged by Eve with $N = 100$ ($N_c = 300$) (above) and $N = 500$ ($N_c = 600$) (below).}
%%\BTdo{Enlarge fonts axis and legends. Remove the title in the top. Enlarge blue points} \MB{The subscripts in the plot should be uppercase. Explain the meaning of red and blues points (clarify the legend). What does $J_{195}$ mean ?}}
%%of the idea behind Theorem \ref{theo_attack_H0_H1}. The behavior of the attack channel $A^*_{\Delta}$ is illustrated.}
%\label{fig.triangle_test}
%\end{figure}
%
%
%
%%%%%%%%%%%%%%%%%%%%%%%%%%%%% single trianlge test %%%%%%%%%%%%%%%%%%%%%%%%
%We denote by $J$ a generic test image (which can be a forgery or not). For notational simplicity, let $d_{J,i} = c_{I_{i}, J} - \lambda \hat{c}_{I_{i}, J} - \eta $, which can be regarded to as a random variable over different images $I_i$.
%%
%The triangle test permits to determine if image $I_i$ has
%been used by Eve to forge $J$. Specifically, for every candidate image $I_i$ in  $\mathcal{C}_{1,pub}$, we have the following composite hypothesis test:
%%
%\begin{equation}
%\label{triangle_test}
%\begin{array}{l}
%H_0: d_{J,i}  \sim f_{J}\\
%H_1: d_{J,i}  \nsim f_{J}.
%\end{array}
%\end{equation}
%%
%%Let  $\mu_{d_{J',i}} = \mu$ and $\sigma_{d_{J',i}} = \sigma$
%%
%%%%%%%%%%%%%%%%%%%%%%% fine test singolo %%%%%%%%%%%%%%%%%
%

For notational simplicity, in the following, given a test image $J$ and a candidate image $I_i$, we let $d_{J,i} = c_{I_{i}, J} - \lambda \hat{c}_{I_{i}, J} - \eta $. In \cite{Goljan} it is shown that the distribution of $d_{J,i}$ is approximately constant with $I_i$ (and $\hat{c}_{I_i, J}$), so we can write:
\begin{equation}
Pr\{d_{J,i}  = x |\hat{c}_{I_i, J}\} \approx f_{J}(x),
\end{equation}
for some $f_{J}$, independent of $I_i$ and $\hat{c}_{I, J}$. Let, $\mu_{J}$ and $\sigma_{J}$ denote the mean and variance of $d_{J,i}$ when $I_i$ is not used by Eve to create the forgery $J$\footnote{This may either correspond to a situation in which $J$ is a forgery but $I_i$ has not been used to create it, or to a case in which $J$ is not a forgery.} (expectedly, $\mu_{J}$ is very close to 0).
In \cite{Goljan}, it is argued that $f_{J}$ is often close to a Gaussian distribution, that is $f_{J} \sim \mathcal{N}(\mu_{J},\sigma_{J})$, even if for some images a Student's $t$-distribution may be a more conservative choice. For sake of brevity, in the following, we stick to the Gaussian model, the difference with respect to the Student's $t$-model being very small based on our experiments.
%
%
%
%%%%%%%%%%%%%%%%%%%%%%%%%%%%%%%%%%
%\BT{************************************}
%The optimum decision for the test in \eqref{triangle_test} cannot be derived since the distribution of  $d_{J,i}$ under $H_1$, that is when $I_i$ is used for the forgery, is not known to Alice and, moreover, it may depend on $I_i$. We generally denote with $\mu_{J,i}^1$ and $\sigma_{J,i}^1$ the mean and the variance of $d_{J,i}$ under $H_1$.
%%Reasonably/approximately, $\mu_{J',i}^1/\sigma_{J',i}^1 > \mu/\sigma$, $\forall i$, we can assume that \eqref{triangle_test} is a {\em one-tailed} test.
%In \cite{Goljan}, the decision threshold $t$ for the triangle test is set by imposing the constraint on the false alarm, that is
%%(to set the decision threshold $t$, the constraint on the false alarm is imposed)
%$Pr\{d_{J,i} > t|H_0\} = P_{fa}$, for a desired $P_{fa}$.
%%
%In doing so, the {\em one-tailed} assumption is made for the test in  \eqref{triangle_test}, which holds true since we can assume that $\mu_{J,i}^1/\sigma_{J,i}^1 > \mu_J/\sigma_J$, $\forall i$.
%
%%*************
%%In imposing this constraint on the false positive, the one-tailed assumption is made. This corresponds to assume that $\mu/\sigma < \mu_{X_i|H_1}/\sigma_{X_i|H_i}$, which is a reasonable assumption (true with high probability). Also, reasonably $\mu_{X_i} > \mu$.
%%We also reasonably assume that $(\mu_{X_i|H_1} - \mu)/\sigma$??????...\\

\enlargethispage{\baselineskip}

\subsection{The pooled triangle test}
%

%The single-image version of the triangle test relies on $d_{J,i}$ to determine if any single image in $\mathcal{C}_{1,pub}$ has been used by Eve to forge $J$. When Alice aims only at determining if $J$ is the result of a PRNU-copy attack based on the public set $\mathcal{C}_{1,pub}$, she can resort to the {\em pooled} version of the test \cite{Goljan}, described below.

Let $J$ be the to-be-tested image and let $H_0$ be the hypothesis that $J$ has not been forged, or, equivalently in our scenario, that no image in $\mathcal{C}_{1,pub}$ has been used by Eve to forge $J$. Let $H_1$ be the opposite hypothesis that some of the images in $\mathcal{C}_{1,pub}$ have been used to forge $J$. Let $k$ be the number of candidate images  considered by Alice to carry out the test (we have $k = N_c$ when the entire public set is used for the test). We denote with $\mathcal{C}_{1,pub}^k$ the corresponding subset. The pooled triangle test described in \cite{Goljan} uses the following statistic to decide if some of the images in $\mathcal{C}_{1,pub}^k$ have been used to forge $J$:
\begin{equation}
\label{pooled_test}
L_{k}^{J} = \sum_{i \in \mathcal{C}_{1,pub}^k} \log\left(f_{J}(d_{J,i})\right).
\end{equation}
When $f_{J}$ is a Gaussian, testing $L_k^{J}$  is very similar in spirit to base the test on the sum of the squared distances. In fact, in such a case, we have
\begin{equation}
\label{pooled_test_Gauss}
L_{k}^{J} = - k \log(\sqrt{2 \pi\sigma^2}) - \sum_{i \in \mathcal{C}_{1,pub}^k} \left(\frac{d_{J,i} - \mu_{J}}{\sqrt{2}\sigma_J}\right)^2.
\end{equation}
By observing that $L_{k}^{J}$ corresponds to the log-likelyhood of the deviations $d_{J,i}$ under $H_0$, the image $J$ is said to be a forgery if $L_{k}^{J} < T$, where $T$ is set by fixing the false alarm probability.

%For a fixed $J$, and large enough $k$, we can invoke the central limit theorem (CLT) and assimilate the distribution of  $L_{k}^{J}$ under $H_0$ to a Gaussian. This allows to fix the detection threshold theoretically, by imposing that the false alarm is lower than a predefined value.

\section{An improved statistic for the pooled test}
\label{sec.proposed_test}

A limit of a test based on $L_k^J$ is that such a statistic considers (the log of) the probability of observing the deviations $d_{J,i}$'s under $H_0$ without exploiting the knowledge we have about the distribution of $d_{J,i}$ under $H_1$. In fact, even if the exact distribution of $d_{J,i}$ under $H_1$ is not known, we know that when the image $I_i$ has been used to forge $J$, the measured correlation $c_{I,J}$ tends to be larger than expected, hence resulting in a larger, positive, value of $d_{J,i}$. More precisely, by assuming (w.l.o.g.) that  ${\mu}_{J}$ is 0, we know that (see also Figure \ref{fig.triangle_test}):
\begin{align}
%Pr\{d_{J,i}  < 0 | \text{ \em \footnotesize $I_i$ has been used to forge $J$}\}  < \nonumber \\
% Pr\{d_{J,i} < 0| \text{ \em \footnotesize $I_i$  has not been used to forge $J$}\}.
Pr\{d_{J,i}  < 0 | \text{ \em \footnotesize $I_i$ used to forge $J$}\}  <
 Pr\{d_{J,i} < 0| \text{ \em \footnotesize $I_i$  not used}\}.
 \end{align}
This is the typical example of one-tailed statistical test, for which the sign of the deviation from the expected value should be taken into account in addition to the magnitude of the deviation.
%
%To introduce our statistics, we start by making the following observation: for any candidate image $I_i$,  when the image $I_i$ is used to forge $J$, we have that, not only the distance of the true correlation from the inference line is large, but also, that is very unlikely that the true correlation  stays below the inference line (see the example in Figure \ref{fig.triangle_test}). To simplify the argument, assume (w.l.o.g.) that  ${\mu}_{J}$, i.e., the mean value of $f_J(d_{J,i})$ for non used $I_i$, is exactly 0 (generally, ${\mu}_{J} \approx 0$). Formally,  what we observe is that
%$$Pr\{d_{J,i}  < 0 | I_i \text{ \em used}\}  \ll Pr\{d_{J,i} < 0| I_i \text{ \em not used}\}$$
%where $Pr\{d_{J,i} < 0 |I_i \text{ \em not used}\} \approx \int_{-\infty}^0 f_J(x) dx$, independent of $i$.
%
Such one-tailed nature of the test is discarded with the  statistic in \eqref{pooled_test_Gauss}, which, by looking at the quadratic distances $d_{J,i}$, implicitly assumes that a large positive and a large negative value of $d_{J,i}$  are equally probable when $I_i$ is used by Eve for the PRNU-copy attack. Note that, even if we exemplified this problem by assuming a Gaussian distribution for $d_{J,i}$, the above observations are generally true for any distribution $f_J$.
Based on the above observation, we propose to replace $L_{k}^{J}$ with a new statistic that takes into account the sign of the deviation $d_{J,i}$, with the understanding that only positive values contribute to form the evidence that $J$ has been forged by Eve. Specifically, we suggest to replace $L_{k}^{J}$ with the following:
\begin{equation}
\label{statistic_VIPP1}
%V_{N_c} = \sum_{i \in N_c} \text{sign}(d_i - \mu) \left(\frac{d_i - \mu}{\sigma}\right)^2,
%V_{K}^{J'} = \sum_{i=1}^K
V_{k}^{J} = \sum_{i \in \mathcal{C}_{1,pub}^k} \text{sign}(d_{J,i} - \mu_{J}) \left(\frac{d_{J,i} - \mu_{J}}{\sigma_{J}}\right)^2,
\end{equation}
%
%\BT{With respect to Goljan, what we consider (or, better, do not neglect), is the one-tailed nature of the triangle test (the one-tailed nature of the pooled is considered by him as well!)}
%
where, as before, $\mu_{J}$ and $\sigma_{J}$ are the mean and variance of $d_{J,i}$
under the hypothesis that $I_i$ has not been used by Eve to forge $J$\footnote{Following \cite{Goljan}, the pooled test is implemented by replacing $\mu_{J}$ and $\sigma_{J}$ with their sample estimates.}.
With reference to \eqref{pooled_test_Gauss}, it is evident that the main difference between $L_k^{J}$ and $V_k^{J}$ is the dependence of $V_k^{J}$ on the sign of $d_{J,i} - \mu_{J}$.
%Basically, the sum of the {\em signed} distances from the line are considered in \eqref{statistic_VIPP1}.
%
In this way, $V_k^{J}$ exploits the knowledge that $Pr\{d_{J,i}  <{\mu}_{J} |H_0\}  > Pr\{d_{J,i} < {\mu}_{J}|H_1\}$, thus resulting in a more accurate test. An additional advantage of directly considering the distances from the inference line rather than the probability values, is that we do not need to make any assumption on the distribution of $d_{J,i}$ for the images not used by Eve ($f_{J}$).
In general, other $n$-powers could be considered for the distance term $(d_{J,i} - \mu_{J})/\sigma_{J}$ in \eqref{statistic_VIPP1}. For instance, we run some experiments by accumulating linear rather than quadratic distances obtaining similar results. In this paper, we chose the square distances to ease the comparison with the statistic $L_k^J$, which in fact results in the accumulation of quadratic distances when $f_J$ is a Gaussian (see \eqref{pooled_test_Gauss}).
%We also experimentally checked that, for some test images $J$,
%sometimes the linear sum ($n = 1$) gives (slightly) better results.
%Further investigation can be carried out in this respect as a future work.
%
%
%Since under $H_0$, $d_i$ are i.i.d., and $N_c$ is generally large, by the central limit theorem, $V_{N_c}$ is approximately large, then, we have the following composite hypothesis test on $Z$ \BT{[We cannot do the same under $H_1$ since the means and the variances may change]}
%%
%%
%\begin{equation}
%\begin{array}{l}
%H_0: Z \sim \mathcal{N}(N_c\mu, \sqrt{N_c}\sigma)\\
%H_1: \mu_Z > N_c\mu
%\end{array}
%\end{equation
%
%%%%%%%%%%%%%%%%%%%%%%%%%%%%%%%%% THEORETICAL MOTIVATION %%%%%%%%%%%%%%%%%%%%%%%%%%%%%%%%%
%Some insights on the statistical motivation for (Deeper insights on) the improvement obtained using $V_{K}^{J'}$ with respect to $L_{K}^{J'}$ can be found in the Appendix \BT{[DECIDE AT THE END IF INCLUDE IT..]}.
%%%%%%%%%%%%%%%%%%%%%%%%%%%%%%%%%%%%%%%%%%%%%%%%%%%%%%%%%%%%%%%%%%%%%%%%%%%%%%%%%%%%%%%%

%a one tailed test can be performed

Eventually, the test decides in favour of $H_1$ if $V_{k}^{J} > T'$, where the threshold $T'$ is fixed by imposing a constraint on the false alarm probability. On this regard, we observe that, as for $L_{k}^{J}$, there are two sources of randomness in $V_{k}^{J}$, namely $J$ and $\mathcal{C}_{1,pub}^k$\footnote{Strictly speaking, $L$ and $V$ depend on the set $\mathcal{C}_{1,pub}^k$. With a slight abuse of notation, we simply denote such a dependence with the letter $k$ in the pedex.}. Then, the false alarm probability can be evaluated by varying either $\mathcal{C}_{1,pub}^k$ or $J$.
In the former case (which is the approach followed in \cite{Goljan} to test the performance of $L_{k}^{J}$), $J$ is fixed, and the distribution of $V_{k}^{J}$ under $H_0$ can be theoretically approximated to a Gaussian.
The terms of the sum in \eqref{statistic_VIPP1}, in fact, are independent under $H_0$, although they are not identically distributed because of the presence of the sign. The central limit theorem can then be applied (the Lindeberg condition \cite{Bill86} is satisfied), and $V_{k}^{J}$ assumed to be normally distributed, thus allowing to set the threshold $T'$ theoretically.

%It is worth saying that considering the quadratic distances (2-power) in \eqref{statistic_VIPP1} is a choice which is expected to expand the differences between the distance values under $H_0^p$ and $H_1^p$. Others $n$-powers could be used getting similar performance.
%%We also experimentally checked that, for some test images $J$,
%%sometimes the linear sum ($n = 1$) gives (slightly) better results.
%Further investigation can be carried out in this respect as a future work.

%Considering the quadratic distances (2-power) \eqref{statistics_VIPP1} is just a choice which is expected to expand the differences between the distance values under $H_0$ and $H_1$. Others powers could be used in principle getting similar performance. We experimentally see that, depending on the test image $J$....,  sometimes the linear sum gives (slightly) better result. Hence, in our experiments we also consider the statistic:
%%
%%
%\begin{equation}
%V_{K}^1 = \sum_{i= 1}^{K} \left(\frac{d_{J',i} - \mu_{J'}}{\sigma_{J'}}\right).
%\end{equation}
%%
%%\BT{Dire che questo e' possibile perche' il guadagno maggiore viene dallo one-tail e non dal quadrato. Dire che come lavoro futuro si potrebbe investigare se vi sia un $n$ migliore di altri (sia $<1$ che $>1$)...MA LIMITARCI AL CASO QUADRATICO QUI}

\section{Experiments}
\label{sec.exp}

%\BT{Is the improvement significant in practice? For that, we need
%the experiments...}
%
%\BT{Describe the setup for your experiments (a Figure may help), the methodology we followed (basically, the two - conceptually different - ways we follow to test the performance pooled detector) and provide the results.}

We run our tests by considering the Nikon D7000 camera ($\textrm{C}_1$) and the Nikon D90 camera ($\textrm{C}_2$) in the RAISE dataset \cite{RAISE8K}. We split the images from $\textrm{C}_1$ as follows: a total number of 1000 images were used to build the public set $\mathcal{C}_{1,pub}$ (in some experiments only a subset of 600 images was used as $\mathcal{C}_{1,pub}$); 300 images were used to build the private set $\mathcal{C}^{(1)}_{1,priv}$, used by Alice to estimate the parameters of the triangle test, that is, to estimate $\lambda$ and $\eta$ and build the inference line; another set $\mathcal{C}^{(2)}_{1,priv}$ of 300 images was used to establish the decision threshold of the correlation detector (with a true positive rate set to 0.9). Other 300 images, passing the correlation test, formed a third set $\mathcal{C}_{1,priv}^{(3)}$ used in the experiments to simulate $H_0$. Eventually, all the 100  flat-field images available in the RAISE dataset for the camera $\textrm{C}_1$ were used to estimate the PRNU.
A number of 300 images coming from a
camera Nikon D90 were used to build Eve's set $\mathcal{C}_{2}$.
The original sizes of the images from Eve's and Alice's cameras $\textrm{C}_1$ and $\textrm{C}_2$ are different. In our experiments, we considered image sizes of $1936\times 1296$ (medium size) and $1024\times1024$ (small size) pixels, obtained by cropping the central parts of the images from  $\textrm{C}_1$ and $\textrm{C}_2$. %In the following we refer to such images as medium and small size images respectively.
With regard to the fingerprint-copy attack performed by Eve, for simplicity, we considered the minimum strength $\alpha$ resulting in a positive identification in the correlation test. This is a worst case assumption for Alice, since in practice Eve can not reproduce exactly Alice's test, and then
she will apply an $\alpha$ which is larger than such a minimum value to be sure to pass the test.

%Let $N$ be the number of images used by Eve and $N_c$ be the total number of images available in $\mathcal{C}_{1,pub}$.
%
We run our experiments by considering two slightly different versions of the pooled test, corresponding to two different interpretations of the error probability and, in particular, the false alarm probability. The two resulting settings correspond to the following testing conditions:
\begin{itemize}
\item[a)]  Given a test image $J$, the error probabilities are computed by varying the subset of $k$ images used to compute $V_k^J$ (res. $L_k^J$). In this setting, the false alarm probability corresponds to the probability that, given $J$, $k$ images at random taken from $\mathcal{C}_{1,pub}$ result in a value of $V_k^J$ (res. $L_k^J$) larger (res. lower), than the detection threshold;
\item[b)] Given $k$ images in $\mathcal{C}_{1,pub}$, the error probabilities are computed by varying the to-be-tested image $J$. In particular, the false alarm probability corresponds to the probability that $\textrm{C}_1$ produces an image for which $V_k^J$ (res. $L_k^J$) is larger (res. lower), than the detection threshold.
\end{itemize}
Two considerations are in order. The setup a) is equal to the one used in \cite{Goljan}. As we have already noticed, in this case both $V_k^J$ and $L_k^J$ can be assumed to be normally distributed, hence the detection threshold can be determined theoretically by fixing the false alarm probability and estimating the mean and variance of the test statistic by resorting to bootstrapping (as in \cite{Goljan}). With regard to b), the distribution of the statistics $V_k^J$ and $L_k^J$ under $H_0$ is not known, so it is not possible to set the detection threshold theoretically by fixing the false alarm probability. In this case, then, we evaluated the performance of the test by plotting the ROC curve of the test any evaluating the missed detection probability for a given false alarm probability set by choosing a suitable operating point on the ROC curve.

\enlargethispage{\baselineskip}

\subsection{Performance of the test for the setup a)}
\label{sec.exp_fix_J}

To test the performance in this case, we fixed the forged image $J$, obtained by taking an image in $\mathcal{C}_2$ and applying the attack in \eqref{forging}.
Then, we picked a random set of $k$ images out of the $N_c$ images in $\mathcal{C}_{1,pub}$, and we computed the statistics $V_k^J$ and $L_k^J$. We repeated this procedure by changing the random selection of the $k$ images, thus getting a number of observations for both statistics under $H_1$.
Finally, we measured the correct detection probability $P_d$, for a fixed theoretical target $P_{fa}$.
Specifically, we computed the $p$-value corresponding to the observed statistics and the image $J$ is said to be forged if the $p$-value of the observation is lower than $P_{fa}$. From the discussion in the previous section, the $p$-value is computed by considering the Gaussian model for $V_{k}^{J}$ (res. $L_{k}^{J}$) under $H_0$.
As in \cite{Goljan},  we let $k = 60$, then we evaluated $P_d$ by bootstrapping, i.e., by repeating the process 30000 times, each time changing the random selection of $k$ images in $\mathcal{C}_{1,pub}$.
%
%\begin{figure}
%\centering
%\subfigure{{\includegraphics[width=0.49\columnwidth]{plots/Boot_P_value_195_2.png}}\label{subfig:original}}
%\subfigure{{\includegraphics[width=0.49\columnwidth]{plots/Boot_P_value_166_2.png}}\label{subfig:attacked}}\\
%%\quad \subfigure[Remapped]{{\includegraphics[width=0.48\columnwidth]{prova.jpg}}\label{subfig:remapped}}\\
%%\vspace*{1cm}
%\subfigure{{\includegraphics[width=0.49\columnwidth]{plots/Boot_P_value_142_2.png}}\label{subfig:Hattacked}}
%\subfigure{{\includegraphics[width=0.49\columnwidth]{plots/Boot_P_value_225_2.png}}\label{subfig:Hmatching}}
%%\quad  \subfigure[Hist. of remapped image]{{\includegraphics[width=0.48\columnwidth]{prova.jpg}}\label{subfig:Hremapped}}
%\caption{$P_d$ as a function of $N$ for 4 images in $\mathcal{C}_2$; $P_{fa} = 10^{-3}$, $N_c = 600$. \BTdo{1. Enlarge fonts! The first point is at $N = 5$. Axis should start from that point! 2. Apex $J$ is missing in the labels. 3. Use different marks!!!!}}
%\label{fig:bootstrapp_pvalue}
%\end{figure}
%
\begin{figure}
\centering
\subfigure{{\includegraphics[width=0.49\columnwidth]{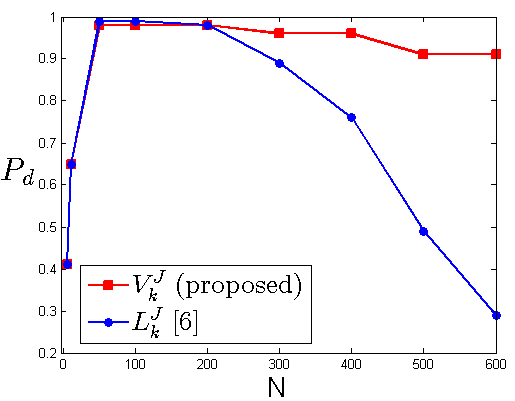}}\label{subfig:original}}
\subfigure{{\includegraphics[width=0.49\columnwidth]{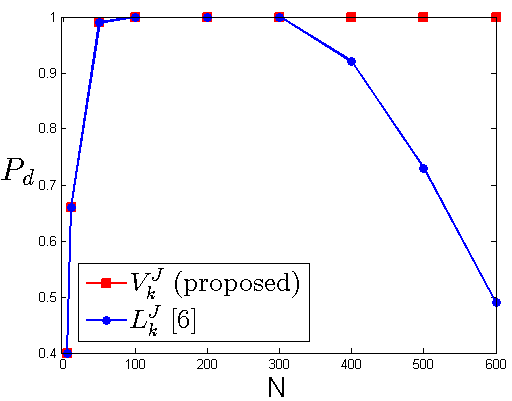}}\label{subfig:attacked}}\\
\caption{$P_d$ as a function of $N$ for 2 images in $\mathcal{C}_2$; $P_{fa} = 10^{-3}$, $N_c = 600$. The minimum $N$ considered is $4$.}
\label{fig:bootstrapp_pvalue}
\end{figure}
Figure \ref{fig:bootstrapp_pvalue} shows the results of the tests
%in terms of $P_d$, when $P_{fa} = 10^{-3}$,
carried out on 2 randomly chosen images in $\mathcal{C}_{2}$. The tests were run for various values of $N$, with $N_c = 600$, and $P_{fa} = 10^{-3}$. For each $N$, the to-be-implanted PRNU $\hat{K}_E$ is estimated from $N$ randomly chosen images in the candidate set. The size of the images is $1936\times 1296$.
We can see that the use of the improved statistic $V_k^J$ brings a significant advantage when $N/N_c > 0.5$, while for small values of the ratio $N/N_c$, the new and the old statistics behave similarly.
%Though not reported here for sake of brevity, we observed a similar behaviour for different values of $N_c$.
A similar behaviour is observed for different values of $N_c$.
In general, the difference between $V_k^J$ and $L_k^J$ can be better appreciated when $N_c$ is large (say $N_c > 300$), since when $N_c$ is small  the pooled test is very powerful and both statistics works very well.

\subsection{Performance of the test for the setup b)}
\label{sec.exp_fix_K}

In this case, we fixed $\mathcal{C}_{1,pub}^k$ and run the pooled test by varying the test image $J$.
We computed the statistics $V_{k}^{J}$ and $L_{k}^{J}$ by forging the images in $\mathcal{C}_2$, whereas the values under $H_0$ were obtained by considering the images in $\mathcal{C}_{1,priv}^3$. Throughout these these experiments we let $k = N_c$. This is a reasonable assumption that corresponds to assuming that Alice knows the entire public set available to Eve.

\begin{figure}
\centering
\subfigure{{\includegraphics[width=0.49\columnwidth]{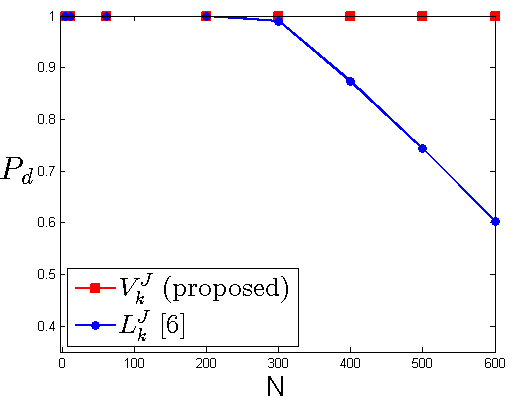}}\label{subfig:original}}
\subfigure{{\includegraphics[width=0.49\columnwidth]{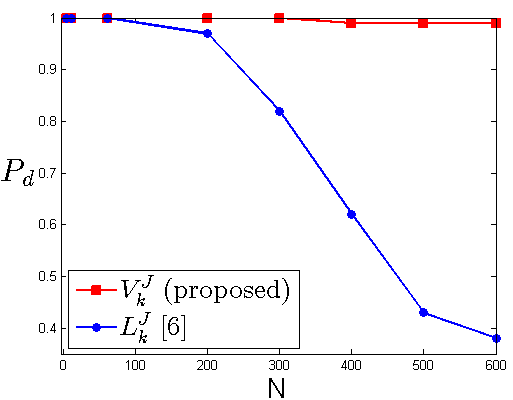}}\label{subfig:attacked}}\\
\caption{$P_d$ values obtained from the ROC curve by letting $P_{fa} = 0.03$, $N_c = 600$. Image size: $1936\times 1296$ (left) and $1024\times1024$ (right). The minimum $N$ considered is $N = 4$.}
%\caption{$P_d$ values obtained by letting $P_{fa} = 0.03$ and fixing the detection threshold based on the ROC curves. Image size: $1936\times 1296$ (left) and $1024\times1024$ (right). Plots start at $N = 4$.}
\label{fig:Pd_from ROC}
\end{figure}

\begin{figure}
\centering
\includegraphics[width=0.5\columnwidth]{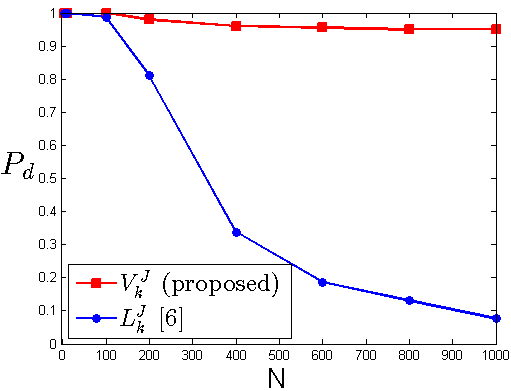}
\caption{$P_d$ values obtained from the ROC curve by letting $P_{fa} = 0.03$, $N_c = 1000$. Image size: $1024\times1024$.}
\label{fig:Pd_from ROC_1000}
\end{figure}

The values of $P_d$ obtained from the ROC curve by fixing the false alarm probability to 0.03
are reported in Figure \ref{fig:Pd_from ROC} for various values of $N$ ($N_c = 600$), for both small and medium size images. The advantage of the improved statistic increases with $N$.
Expectedly, with small images the performance of the pooled test are lower and the difference between the two statistics is more evident.
%when $N = 600$, we have $P_d = 0.61$ for $L_k^J$ and $P_d = 1$ for $V_k^J$ for medium size images, and $P_d = 0.46$ ($L_{k}^{J}$) and $P_d = 0.99$ ($V_{k}^{J}$), with the small images.
We observe that the test achieves perfect results also when $N/N_c$ is very low.
This is a consequence of the fact that $k = N_c$ (or, more in general, that $k$ is comparable to $N_c$), since with this choice the pooled test is very reliable especially when $N/N_c$ is small.
A similar behaviour holds for other values of $N_c$. Figure \ref{fig:Pd_from ROC_1000} shows the results we have got with $N_c = 1000$ in the least favorable case of small size images. We see that the test with  $V_k^J$ is still reliable with such a large $N_c$: in particular, at $N = 1000$, we get  $P_d = 0.95$ , while, for the test with $L_k^J$, $P_d$ is $0.0767$. We verified that for the case of medium size images  we still get very close-to-ideal performance with  $N_c = 1000$ (in the most difficult case with $N = 1000$, we get $P_d = 0.99$ with $V_k^J$, and $P_d = 0.15$ with $L_k^J$).

%%%%%
%
%\begin{table}
%%
%\begin{center}
%\caption{$P_d$ values at $P_{fa} = 0.03$ derived from the ROC curves for both image sizes.\BT{N =100 missing}}
%\small
%%\renewcommand\arraystretch{1.3}
%\setlength{\tabcolsep}{3pt}
%\vspace{-0.2cm}
%\begin{tabular}{c   c  |c c c c c c c c c}
%    %\multirow{2}{*}{}   \multirow{2}{*}{}         &         \multicolumn{9}{c}{\bf N}   \\
%        \multicolumn{2}{c}{}       &         \multicolumn{9}{c}{N}   \\
%           &   & 4 & 10 & 60 & 100 & 200 & 300 & 400 & 500 & 600 \\ \hline
%\multirow{2}{*}{{\small $1936 \times 1296$}} &  [6] & 1 & 1 & 1 & ? & 1 & 0.99 & 0.90 & 0.76 & 0.61 \\
%&  (our) & 1 & 1 & 1 & ? & 1 & 1 & 1 & 0.99 & 0.99 \\ \hline
%\multirow{2}{*}{{\small $1024\times 1024$}} &  [6] & 1 & 1 & 1 & ? & 0.97 & 0.83 & 0.68 & 0.49 & 0.46 \\
%&  (our) & 1 & 1 & 1 & ? & 1 & 1 & 0.99 & 0.99 & 0.99 \\ \hline
%\end{tabular}
%\label{tab:Pd_from ROC}
%\end{center}
%\normalsize
%%\vspace{0.3cm}
%\end{table}

%
%These results show that,the pooled test, especially the version proposed in this paper, is very powerful.
%In particular, from our experiments, we argue that, using  $V_K^J$, a reliable test can be carried out even  when $N$ is much larger than the value of few hundred, regarded to as a kind of limit value for reliable detection in \cite{Goljan}. \BT{[Jessica stopped her analysis at N =300. From the paper, that value is looked at as a kind of limit $N$ for reliable detection]
%We should be careful since we cannot compare our results with those of Jessica (so we cannot report the 300 number). There are some differences, not least the fact that Jessica considers JPEG images.}\\
%\BTcomm{We should include at least a case with a larger $N_c$ in the paper. But do we want to investigate such a limit for $N$?  Moreover, it may also depend on $N_c$.... \\Based on these tests we are doing in these days, $N = N_c = 1000$ is not sufficient.....it seems a bit too much, but it is not totally unreasonable.....We are raising it further (these tests can be time consuming and are not our main goal).}

\section{Conclusions}
\label{sec.conc}

We have proposed a new statistic for the pooled triangle test originally introduced in \cite{Goljan}. The improved statistic is based on the observation that the statistic proposed in \cite{Goljan} somewhat neglects the one-tailed nature of the test. Experiments show that the proposed statistic achieves better results, especially in the most challenging case when the number of images $N$ used by Eve for the fingerprint-copy attack is large (and comparable to $N_c$).
Further tests could be carried out to investigate the limit values of $N$ (and $N_c$) for which the test based on the new statistic is still reliable. As a further work, we plan to evaluate the performance of the pooled test based on the improved statistic in the presence of targeted attacks like those introduced in \cite{rao2013anti, Caldelli}.

\section*{Acknowledgements}

We thank M. Goljan and J. Fridrich from Binghamton University for their help in clarifying some details of the pooled triangle test. H{\'e}ctor Santoyo Garc{\'i}a thanks the National Council of Science and Technology (CONACyT) of Mexico for financial support and Prof. Mariko Nakano-Miyatake for financial support and advice.

\bibliographystyle{IEEEtran}
\bibliography{SPLbiblio}

%\appendix
%
%\BTcomm{I have left it out the insights on the statistical motivation. There is no room anyhow.}
%%\BTcomm{If there is not enough space as I fear, we will for sure not include it :-), then I did not waste time adjusting notation ...}

\end{document}